\documentclass[preprint,floatfix] {revtex4} 
\newcommand{\rvec}{\mathrm {\mathbf {r}}} 
\usepackage{graphicx}
\usepackage{subfigure}
\begin{document}

\title{Pseudopotential density functional treatment of atoms and molecules in cartesian coordinate grid}
\author{Amlan K.\ Roy}
\altaffiliation{Present address: Indian Institute of Science Education and Research (IISER), Division of Chemical Sciences, 
Block HC, Sector III, Salt Lake City, Kolkata-700106.}
\affiliation{Department of Chemistry, University of Kansas, Lawrence, KS, 66045, USA.}
\email{akroy@chem.ucla.edu, akroy@iiserkol.ac.in}

\begin{abstract}
This is a follow-up of our recently proposed work on pseudopotential calculation (Ref. [21]) of atoms and molecules within 
DFT framework, using cartesian coordinate grid. Detailed results are presented to 
demonstrate the usefulness, applicability of the same for a larger set of species (5 atoms; 53 molecules) and 
exchange-correlation functionals (local, nonlocal). A thorough comparison on total, component, ionization, atomization
energies, eigenvalues, potential energy curves with available literature data shows excellent agreement. 
Additionally, HOMO energies for a series of molecules show significant improvements by using the Leeuwen-Baerends 
exchange potential, compared to other functionals considered. Comparison with experiments has been made, wherever 
possible.

\end{abstract}
\maketitle

\section{Introduction}
Development of accurate efficient methods for the electronic structure of molecules, solids, clusters has been a topic of 
increasingly intense interest in chemistry, physics, materials science, etc. In the past two decades, density functional 
theory (DFT) has been applied with remarkable success to investigate the structural (such as geometries, vibrational 
frequencies, energetics of chemical reactions, etc.) as well as dynamical (such as photoabsorption, photoemission, 
multi-photon
ionization, high-order harmonic generation, etc.) characteristics of such systems. In essence, the ground-state electronic 
energy is partitioned as: $E[\rho(\rvec)] = T_s[\rho(\rvec)] + U[\rho(\rvec)] + E_{xc}[\rho(\rvec)]$, where the three 
terms in the right-hand side denote kinetic energy of a system of non-interacting particles of same density $\rho(\rvec)$, 
classical electrostatic energy (including electron-nuclear attraction, electron-electron and nuclear-nuclear repulsion), 
and exchange-correlation (XC) energy respectively. The single-particle wave functions are obtained from the Kohn-Sham (KS)
equation as:
\begin{equation} 
\left[ \frac{1}{2} \nabla^2 + v_{eff}(\rvec) \right] \psi_i(\rvec) = \epsilon_i \psi_i (\rvec)
\end{equation} 
where $v_{eff}(\rvec)$ contains the relevant one- and two-body potentials, while the electron density is given by the 
squared sum of the occupied orbitals, $\rho(\rvec)= \sum_i f_i |\psi_i(\rvec)|^2 $.

Broadly speaking, two approaches have gained popularity for the practical solution of above KS equation. In the 
\emph{real-space method} \cite{white89, chelikowsky94, briggs95, gygi95, beck00} the discretized KS equation is solved 
iteratively on a mesh using either finite-difference, finite-element or wavelets. The grid-based matrix presentation 
produces \emph{structured, highly banded} matrices. Moreover, the near \emph{locality} (the potential operator is diagonal 
in coordinate space whereas the Laplacian operator is nearly local) can potentially remove the omnipresent nagging problem 
arising 
due to the basis-set incompleteness in the contrasting basis-set approaches. Furthermore, they are easily amenable to the 
so-called linear-scaling methods. While such schemes usually require a larger number of grids to achieve physically
meaningful results, through introduction of higher-order and multigrid techniques, the effective grid can be reduced 
significantly. In \emph{basis-set} approaches, on the other hand, the single-particle wave functions are represented by 
a variety of functions such as Slater-type orbitals, gaussian type functions (GTF), numerical functions, plane waves (PW), 
augmented plane waves, linear muffin-tin orbitals etc. While currently, there is a preponderance of the atom-centered 
localized GTFs in the field of molecular quantum chemistry \cite{helgaker00} (chiefly due their ability to deal with some 
of the important multi-center integrals analytically), \emph{ab initio} calculation in the condensed matter regime is 
almost exclusively dominated by the PWs \cite{payne92} (mainly because the Fast-Fourier transform (FFT) technique can be 
employed to take advantage of the periodicity of such systems). This work is concerned with the basis-set approach to tackle
the electron-structure calculation of atoms and molecules. Recently there has been effort to combine the basis sets; 
e.g., in \cite{krack00}, a gaussian basis set was used to expand the KS molecular orbital (MO), whereas an \emph{auxiliary} 
augmented PW basis was utilized to express the electronic charge density. 

Excepting a very few attempts such as the \emph{grid-free} approach \cite{zheng93}, that uses a resolution of identity 
to evaluate multi-center integrals over functionals at the expense of an \emph{auxiliary} basis set, a large majority of 
basis-set-based molecular DFT calculations employ some carefully selected suitable 3D quadrature for a sensible choice of 
grid points in space. A very successful common scheme \cite{becke88} involves partitioning the 3D molecular integrand 
into single-center discrete atomic ``cells". These can be treated using standard numerical techniques individually and then 
summing these up using some appropriate weight functions leads to the desired final result. In this so-called 
atom-centered grid (ACG) method, the mono-centric atomic integrals are computed by separating the radial and angular 
components. The former has typically been performed by introducing several quadratures and various mapping schemes. Some 
of these are Gauss-Chebyshev quadrature of second kind, Chebyshev quadratures of first and second kind, Gaussian quadrature, 
Euler-MacLaurin formula, numerical quadratures, etc. \cite{treutler95, mura96, murray93, gill93, lindh01}. Angular 
integrations are handled usually by the Lebedev spherical method \cite{lebedev75, lebedev76, lebedev92}. In another development, 
\cite{boerrigter88} 3D integration of molecular integrands were performed numerically based on a division of space and
subsequent integration over the resulting regions by product Gauss rule. A variational integration mesh \cite{pederson90}, which
depends on the position of individual atoms has also been reported by breaking the space into three different regions, 
\emph{viz.,} atomic spheres, excluded cubic region and interstitial parallelepiped. Other numerical 
grids \cite{chen95} have been proposed as well for gaussian basis set calculations. Recently there has been an attempt to 
connect the cartesian coordinate grid (CCG) and ACG by a divided-difference polynomial interpolation which can translate 
the electron density and gradients from the former to the latter \cite{kong06}. 

In a previous article \cite{roy08}, henceforth referred to as I, efforts were made to employ the CCGs in the context 
of atoms and molecules within the linear combination of GTF ansatz of DFT. In this work, respective quantities such as 
the localized atom-centered basis set, the two-body classical Coulomb repulsion and the non-classical XC potentials were 
directly set up on the 3D CCG. Local-density approximated (LDA) XC functionals of the homogeneous electron gas in 
conjunction with the Hay-Wadt type pseudopotential was used. A multitude of quantities like \emph{viz.,} individual energy 
contributions to total energy, eigenvalues, potential energy curves, ionization energies, as well as atomization energies 
were investigated. The relevance and performance of the CCGs were judged on 12 molecules (from small to medium size) and 
3 atoms. In each of these cases practically identical results as those from the ACG and grid-free results were 
obtained. Now, it is well-known that the local density functionals suffer from a number of problems and hence it is 
essential to use more accurate functionals, namely those incorporating the gradient and Laplacian corrections. We have a 
number of objectives in this article. Firstly, we want to assess the validity and efficacy of the aforementioned scheme of I 
for a larger set of species to justify and validate its future usage. Secondly its scope of applicability is broadened by 
incorporating the gradient corrected XC functionals (for demonstration, the popular non-local Becke exchange \cite{becke88a} 
and Lee-Yang-Parr (LYP) correlation \cite{lee88} is used). First, a detailed test on the convergence 
of our results on 5 atoms and 12 molecules is made as in I for the BLYP XC functional. In the next step, additional 41
molecules are considered both at the LDA and BLYP level. Detailed comparison with reference theoretical and experimental 
results are made, wherever possible. Besides, we also report the highest-occupied molecular orbital (HOMO) energies
obtained with the Leeuwen-Baerends (LB) \cite{leeuwen94, schipper00} exchange potential for the latter 41 molecules 
(see Section III for motivational details). In all cases, this significantly improves both LDA and BLYP HOMO energies. 
Section II gives a summary of the method of calculation and computational aspects. Section III presents a discussion on 
the results obtained. We end with a few concluding remarks about the future prospects in Section IV.

\section{Methodology and implementation}
This section sketches the essential steps involved in the ground-state calculation of a many-electron system within the 
KS DFT framework used in the present work; further details can be found in \cite{roy08}. 

The KS MOs $\{\psi_i^{\sigma}(\rvec)\}, \sigma=\alpha,\beta$ are linearly expanded in terms of a set of K known basis 
functions as,
\begin{equation}
\psi_i^{\sigma}(\rvec) = \sum_{\mu=1}^K C_{\mu i}^{\sigma} \chi_{\mu} (\rvec), \ \ \ i=1,2,\cdots,K 
\end{equation}
where the set $\{ \chi_{\mu}(\rvec)\}$ denotes the contracted gaussian functions centered on the constituent atoms while 
$\{C_{\mu i}^{\sigma}\}$ contains the contraction coefficients for the orbital $\psi_i^{\sigma}(\rvec)$. The 
individual spin-densities are given by,
\begin{equation}
\rho^{\sigma}(\rvec) = \sum_i^{N^{\sigma}} |\psi_i^{\sigma}(\rvec)|^2 = 
\sum_{\mu} \sum_{\nu} P_{\mu \nu}^{\sigma} \chi_{\mu} (\rvec) \chi_{\nu}^* (\rvec)
\end{equation}
where $P^{\sigma}$ stands for the respective density matrices.

Substitution of these terms into the energy expression, followed by minimization with respect to unknown coefficients 
$C_{\mu i}^{\sigma}$, subject to the orthogonality constraint leads to the following matrix KS equation which is akin to 
the unrestricted Pople-Nesbet equation in Hartree-Fock (HF) theory,
\begin{equation}
\mathbf{F}^{\sigma} \mathbf{C}^{\sigma} = \mathbf{SC}^{\sigma} \mathbf{\epsilon}^{\sigma}.
\end{equation}
Here $\mathbf{S}$ and $\mathbf{F}$ denote the $K \times K$ real, square symmetric overlap and total KS matrices
respectively. $\mathbf{C}$ stands for the eigenvector matrix containing the expansion coefficients $C_{\mu i}^{\sigma}$ 
while the orbital energies $\epsilon_i$ are embedded in the diagonal matrix $\epsilon$. The KS matrix is conveniently
written as,
\begin{equation}
F^{\sigma}_{\mu \nu} = H_{\mu \nu}^{\mathrm{core}} +J_{\mu \nu}+F_{\mu \nu}^{XC\sigma}
\end{equation}
where the first two terms in the right-hand side signify the matrices of one-electron bare-nucleus Hamiltonian and
the classical Coulomb repulsion. The one-electron overlap, kinetic-energy and nuclear-attraction integrals involved are 
the same as encountered in the GTF-based HF schemes and are computed using standard recursion algorithm 
\cite{obara86, mcmurchie78}. The pseudopotential matrix elements in the gaussian basis are taken taken from the GAMESS
quantum chemistry program output \cite{schmidt93}. The Coulomb potential is evaluated using a Fourier convolution 
technique \cite{martyna99, minary02}, which effectively uses a Ewald summation type decomposition in terms of short- and 
long-range contributions,
\begin{equation}
\frac{1}{r}=\frac{\mathrm{erf}(\alpha r)}{r}+ \frac{\mathrm{erfc}(\alpha r)}{r} 
           =v_c^{\mathrm{long}}(r)+v_c^{\mathrm{short}}(r).
\end{equation} 
Here erf(x) and erfc(x) identify the error function and complimentary error function respectively. Note that the latter 
goes to zero exponentially fast at large $r$. The parameter $\alpha$ controls the range on which 
$v_c^{\mathrm{short}}$ is nonzero. After a thorough check on the convergence with respect to $\alpha$, a value within the 
range of 6--8 seemed satisfactory and we employed 7 for all the results reported in this work. This has produced highly 
satisfying results for a modest number of atomic (3) and molecular (12) systems in our previous work \cite{roy08}. The 
short-range Fourier integral is calculated analytically, whereas the long-range part is obtained from FFT of the 
real-space values.

As already mentioned, a major thrust of the current communication is to demonstrate the feasibility and viability of our
current scheme in the context of so-called ``non-local'' (gradient and Laplacian-dependent) XC functionals which would 
be necessary for future chemical applications. For this, we first test this with two of the 
widely used functionals, namely the Becke exchange \cite{becke88a} and LYP correlation \cite{lee88} (for convenience, 
an alternative equivalent form \cite{miehlich89} containing only the first derivative has been mostly used in practice). 
The homogeneous electron-gas correlation of Vosko-Wilk-Nusair (VWN) \cite{vosko80} is used in all the LDA calculations. 
Following \cite{pople92}, the gradient-dependent functionals can be treated without evaluating the density Hessians by 
using a finite-orbital basis expansion. To this end, XC contributions of the KS matrix is written in the following 
working form, 
\begin{equation}
F_{\mu \nu}^{XC \alpha}= \int \left[ \frac{\partial f}{\partial \rho_{\alpha}} \chi_{\mu} \chi_{\nu} +
 \left( 2 \frac{\partial f}{\partial \gamma_{\alpha \alpha}} \nabla \rho_{\alpha} + 
    \frac{\partial f} {\partial \gamma_{\alpha \beta}} \nabla \rho_{\beta} \right) 
    \cdot \nabla (\chi_{\mu} \chi_{\nu}) \right] d\rvec
\end{equation}
where $\gamma_{\alpha \alpha} = |\nabla \rho_{\alpha}|^2$, 
$\gamma_{\alpha \beta} = \nabla \rho_{\alpha} \cdot \nabla \rho_{\beta}$, $\gamma_{\beta \beta} = 
|\nabla \rho_{\beta}|^2$ and $f$ is a function only of the local quantities $\rho_{\alpha}$, $\rho_{\beta}$ and their 
gradients. The BLYP functionals are implemented using the Density Functional Repository program \cite{repository}. The 
two-electron matrix elements cannot be evaluated analytically; here we use direct numerical integrations on the CCG grid. 
Note that some of the existing DFT codes \cite{andzelm92} use an alternative route of fitting these by an auxiliary 
basis set, the so-called discrete variational method \cite{baerends73,sambe75,dunlap79}. The generalized matrix-eigenvalue problem 
is solved using the standard LAPACK routine \cite{lapack99} accurately efficiently. Self-consistent set of MOs, density and energy 
are obtained in the usual manner subject to the convergence of (a) potential (b) total energy and (c) eigenvalues. Tolerance of 
$10^{-6}$ a.u. was used for (b) and (c), while a value of $10^{-5}$ for (a). Atomic units employed, unless otherwise mentioned.
 
\section{Results and discussion}
At first, Table I shows a comparison of our calculated ground-state energy components with respect to the number of mesh 
points $N_r (r \in \{x,y,z\})$ as well as the grid spacing $h_r$, for Cl$_2$ and HCl at internuclear distances 4.2 and 2.4 
a.u respectively employing the BLYP density functional. In this occasion, we adopt a similar presentation strategy as in 
I. A series of calculations were performed in the same fashion to test the convergence of our results with respect
to the grid parameters. These numerical experiments produced very similar conclusions as reached in for LDA XC-case in I. Hence out of eight, we eventually report here the results from two such sets only (to avoid too many entries in the 
table), \emph{viz.,} (i) Set A with $N_r=64$, $h_r=0.3$, (ii) Set B with $N_r=128$, $h_r=0.2$, which suffices to illustrate 
the important points. The reference theoretical results presented in this and all other tables throughout the article imply
those obtained from the GAMESS quantum chemistry program \cite{schmidt93}. They use same basis set, effective core potential and
employing the ``grid" option. The corresponding results from the ``gridfree" option are quoted in footnotes for convenience,
in several occasions. Now onwards, we will refer to them as \emph{grid} and \emph{grid-free} theoretical reference results. The
former uses the Euler-McLaurin quadrature for the radial integrations and Gauss-Legendre quadrature for the angular integrations.
The convergence of energies and other quantities with respect to the radial and angular grid was monitored by performing two 
extra set of calculations (i) $N_r, N_{\theta}, N_{\phi}$ = 96, 36, 72 (ii) $N_r, N_{\theta}, N_{\phi}$ = 128, 36, 72, besides the 
default grid option ($N_r, N_{\theta}, N_{\phi}$ = 96, 12, 24), where the three integers denote the respective number of integration 
points in $r, \theta, \phi$ directions. Note that all the 3 grids offered very similar results; for example, out of the 17
atoms and molecules, total energies remain same upto 5th decimal place for 8 species for all 3 grids. In the remaining cases, 
they differed slightly among each other as the grid parameters changed; the largest deviation in total energy being 0.00064 a.u.
for Na$_2$Cl$_2$ and for all others it is well below 0.00007 a.u. However, passing from the default grid to (ii) gradually 
improves $N$. In this and all other tables in the article, we have quoted (ii) results for the reference \emph{grid}-DFT values.
The \emph{grid-free} implementation uses the resolution of identity to simplify the molecular integrals enabling their analytical
evaluation and obviating the necessity of using grid quadratures. 
Quantities considered are the same as those in I, \emph{viz.,} various energies such as kinetic ($T$), total nucleus-electron 
attraction ($V_t^{ne}$), classical Coulomb repulsion ($E_h$), exchange ($E_x$), correlation ($E_c$), total two-electron 
potential ($V_t^{ee}$), nuclear repulsion ($E_{nu}$), total potential ($V$), electronic ($E_{el}$) and total energy ($E$) 
respectively, as well as the integrated electron density $N$. Evidently, as in the LDA case, both sets offer excellent 
agreement in total energies and component energies with literature values for both molecules. The individual 
two-body energy terms were not available in the reference output and thus could not be directly compared. As expected, for 
obvious reasons Set B results are closer to reference values than Set A, but only marginally. For Cl$_2$ this is slightly 
more pronounced than that for HCl. Absolute deviations in total energy for Set B for Cl$_2$  and HCl are only 0.00002 and 
0.00000 a.u. respectively. In both cases, there is slight improvement in $N$, as we move from Set A to B. For all practical
purposes, Set A is adequate enough for both of them. Note that reference \emph{grid-free} DFT energies differ substantially from 
the corresponding \emph{grid}-DFT values.

\begingroup
\squeezetable
\begin{table}      
\caption{\label{tab:table1} Variation of the energy components and $N$ with respect to the grid parameters for Cl$_2$ and
HCl with reference values. BLYP results in a.u.}
\begin{ruledtabular}
\begin{tabular} {lrrrrrr}
    & \multicolumn{3}{c}{Cl$_2$ ($R=4.2$ a.u.)}  &  \multicolumn{3}{c}{HCl ($R=2.4$ a.u.)}      \\
\cline{2-4} \cline{5-7} 
Set                        &     A      &    B       & Ref.~\cite{schmidt93}     &   A  &   B   &  Ref.~\cite{schmidt93} \\
$N_r$                      &    64      &   128      &             &   64        &    128      &             \\
$h_r$                      &    0.3     &   0.2      &             &   0.3       &    0.2      &             \\
$\langle T \rangle$        & 11.21504   & 11.21577   &  11.21570   & 6.25431     & 6.25464     & 6.25458     \\
$\langle V^{ne}_t \rangle$ & $-$83.72582& $-$83.72695& $-$83.72685 & $-$37.29933 & $-$37.29987 & $-$37.29979 \\
$\langle E_h \rangle$      & 36.74464   & 36.74501   &             & 15.86078    & 15.86103    &             \\
$\langle E_x \rangle$      & $-$5.29009 & $-$5.29015 &             & $-$3.01023  & $-$3.01026  &             \\
$\langle E_c \rangle$      & $-$0.37884 & $-$0.37892 &             & $-$0.21171  & $-$0.21174  &             \\
$\langle V^{ee}_t \rangle$ & 31.07572   & 31.07594   &  31.07594   & 12.63884    & 12.63903    & 12.63901    \\
$\langle E_{nu} \rangle$   &  11.66667  & 11.66667   &  11.66667   & 2.91667     & 2.91667     & 2.91667     \\
$\langle V \rangle$        & $-$40.98344& $-$40.98434& $-$40.98424 & $-$21.74382 & $-$21.74417 & $-$21.74411 \\
$\langle E_{el} \rangle$   & $-$41.43506& $-$41.43524& $-$41.43522 & $-$18.40618 & $-$18.40620 & $-$18.40620 \\
$\langle E \rangle$        & $-$29.76840& $-$29.76857& $-$29.76855\footnotemark[1] & 
                             $-$15.48951 & $-$15.48953 & $-$15.48953\footnotemark[2] \\
 $N$                       & 14.00006   & 14.00000   & 13.99998    & 8.00002     & 8.00000     &  8.00000    \\
\end{tabular}
\end{ruledtabular}
\footnotetext[1]{The \emph{grid-free} DFT value is $-$29.74755 a.u. \cite{schmidt93}.} 
\footnotetext[2]{The \emph{grid-free} DFT value is $-$15.48083 a.u. \cite{schmidt93}.} 
\end{table}
\endgroup

To further demonstrate the accuracy and reliability of present results, in Table II, calculated negative eigenvalues 
for Cl$_2$ and HCl (using BLYP XC combination) are presented at the same geometries of previous table, along with those 
obtained from reference \cite{schmidt93}. Once again, excellent agreement is observed for both molecules. Sets A and 
B results match \emph{completely} with literature values for \emph{all} the orbital energies except the lone case of 
$3\sigma_g$ for Cl$_2$ (Set A), where the absolute deviation is only 0.0001 a.u.
 
\begingroup
\squeezetable
\begin{table}
\caption{\label{tab:table2} Comparison of the calculated negative eigenvalues of Cl$_2$ and HCl with the reference values.
BLYP results are given in a.u.} 
\begin{ruledtabular}
\begin{tabular} {lccclccc}
    MO   & \multicolumn{3}{c}{Cl$_2$ ($R=4.2$ a.u.)} & MO  & \multicolumn{3}{c}{HCl ($R=2.4$ a.u.)} \\
\cline{2-4}  \cline{6-8}
Set          & A      &   B     & Ref. \cite{schmidt93} &  &  A       & B       & Ref. \cite{schmidt93} \\ 
 $2\sigma_g$ & 0.8143 & 0.8143  & 0.8143   & $2\sigma$     &  0.7707  & 0.7707  & 0.7707   \\
 $2\sigma_u$ & 0.7094 & 0.7094  & 0.7094   & $3\sigma$     &  0.4168  & 0.4167  & 0.4167   \\
 $3\sigma_g$ & 0.4170 & 0.4171  & 0.4171   & $1\pi_x$      &  0.2786  & 0.2786  & 0.2786   \\
 $1\pi_{xu}$ & 0.3405 & 0.3405  & 0.3405   & $1\pi_y$      &  0.2786  & 0.2786  & 0.2786   \\
 $1\pi_{yu}$ & 0.3405 & 0.3405  & 0.3405   &  &            &          &                    \\ 
 $1\pi_{xg}$ & 0.2778 & 0.2778  & 0.2778   &  &            &          &                    \\
 $1\pi_{yg}$ & 0.2778 & 0.2778  & 0.2778   &  &            &          &                     \\
\end{tabular}
\end{ruledtabular}
\end{table}
\endgroup

Now Table III tabulates our computed negative total energies of Cl$_2$ (relative to $-$29 a.u. in columns 2--3, left panel) 
and HCl (relative to $-$15 a.u. in columns 6--7, right panel) for Sets A and B using the BLYP XC functionals. A broad range 
of internuclear distance is considered in columns 1 and 5 (3.5--5.0 a.u. for the former and 1.5--3.0 a.u. for the latter) 
and the corresponding \emph{grid} DFT results obtained from the GAMESS program \cite{schmidt93} are listed in columns 5 
and 8 for comparison. These are depicted vividly for smaller ranges of $R$ in Fig.~1. Clearly, for both the molecules, Sets 
A and B results are practically coincident on reference values for the entire range of $R$. For Cl$_2$, maximum absolute 
deviation is 0.0001 a.u. with Set B and 0.0002 a.u. (only in 2 instances) with Set A. However, for HCl, the two corresponding 
maximum deviations are 0.0001 a.u., for both sets. This is anticipated from the results of Table I, where we noticed results 
from these two sets confirmed to each other more for HCl than for Cl$_2$. This discussion clearly illustrates the faithfulness 
of current calculation.
 
\begingroup
\squeezetable
\begin{table}
\caption {\label{tab:table3}Calculated potential energy curves of Cl$_2$ and HCl for grid Sets A, B, along with literature
values (grid-DFT). Negative values are given in a.u.} 
\begin{ruledtabular}
\begin{tabular}{lccclccc}
R (a.u.)   & \multicolumn{3}{c}{Cl$_2$ (Total energy relative to $-$29 a.u.)} & R (a.u.)  & \multicolumn{3}{c}
{HCl (Total energy relative to $-$15 a.u.)}   \\ 
\cline{2-4}  \cline{6-8}
Set   & A       &  B      &  Ref. \cite{schmidt93}  &    & A    & B      & Ref. \cite{schmidt93}  \\ 
3.50  & 0.7032  & 0.7032  &  0.7033  &   1.50   &  0.0533  &  0.0534  &  0.0533  \\
3.60  & 0.7231  & 0.7232  &  0.7231  &   1.60   &  0.1818  &  0.1818  &  0.1818  \\
3.70  & 0.7383  & 0.7384  &  0.7384  &   1.70   &  0.2767  &  0.2767  &  0.2767  \\
3.80  & 0.7497  & 0.7498  &  0.7497  &   1.80   &  0.3464  &  0.3464  &  0.3464  \\
3.90  & 0.7579  & 0.7580  &  0.7579  &   1.90   &  0.3969  &  0.3969  &  0.3969  \\
4.00  & 0.7635  & 0.7635  &  0.7635  &   2.00   &  0.4328  &  0.4328  &  0.4329  \\
4.10  & 0.7668  & 0.7669  &  0.7669  &   2.10   &  0.4578  &  0.4577  &  0.4577  \\
4.20  & 0.7684  & 0.7686  &  0.7685  &   2.20   &  0.4741  &  0.4742  &  0.4742  \\
4.30  & 0.7686  & 0.7688  &  0.7687  &   2.30   &  0.4842  &  0.4842  &  0.4842  \\
4.40  & 0.7676  & 0.7678  &  0.7678  &   2.40   &  0.4895  &  0.4895  &  0.4895  \\
4.50  & 0.7658  & 0.7659  &  0.7659  &   2.50   &  0.4912  &  0.4912  &  0.4912  \\
4.60  & 0.7632  & 0.7633  &  0.7633  &   2.60   &  0.4903  &  0.4903  &  0.4903  \\
4.70  & 0.7599  & 0.7601  &  0.7601  &   2.70   &  0.4874  &  0.4874  &  0.4874  \\
4.80  & 0.7565  & 0.7566  &  0.7566  &   2.80   &  0.4831  &  0.4831  &  0.4831  \\
4.90  & 0.7527  & 0.7528  &  0.7528  &   2.90   &  0.4778  &  0.4778  &  0.4778  \\
5.00  & 0.7486  & 0.7487  &  0.7488  &   3.00   &  0.4718  &  0.4718  &  0.4718  \\
\end{tabular}                                                                               
\end{ruledtabular}
\end{table}
\endgroup

\begin{figure}
\begin{minipage}[c]{0.40\textwidth}
\centering
\includegraphics[scale=0.45]{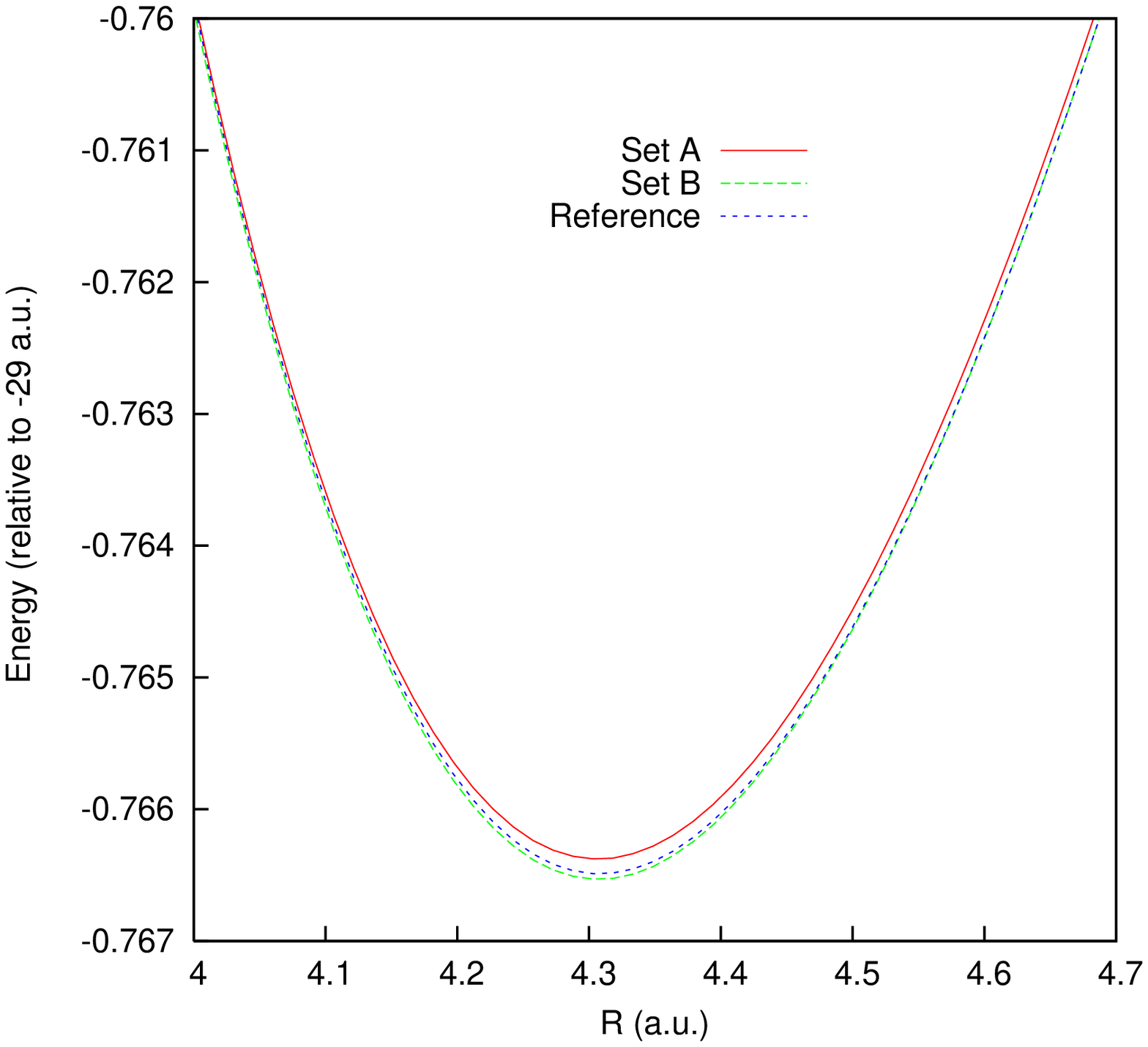}
\end{minipage}%
\hspace{0.8in}
\begin{minipage}[c]{0.40\textwidth}
\centering
 \includegraphics[scale=0.45]{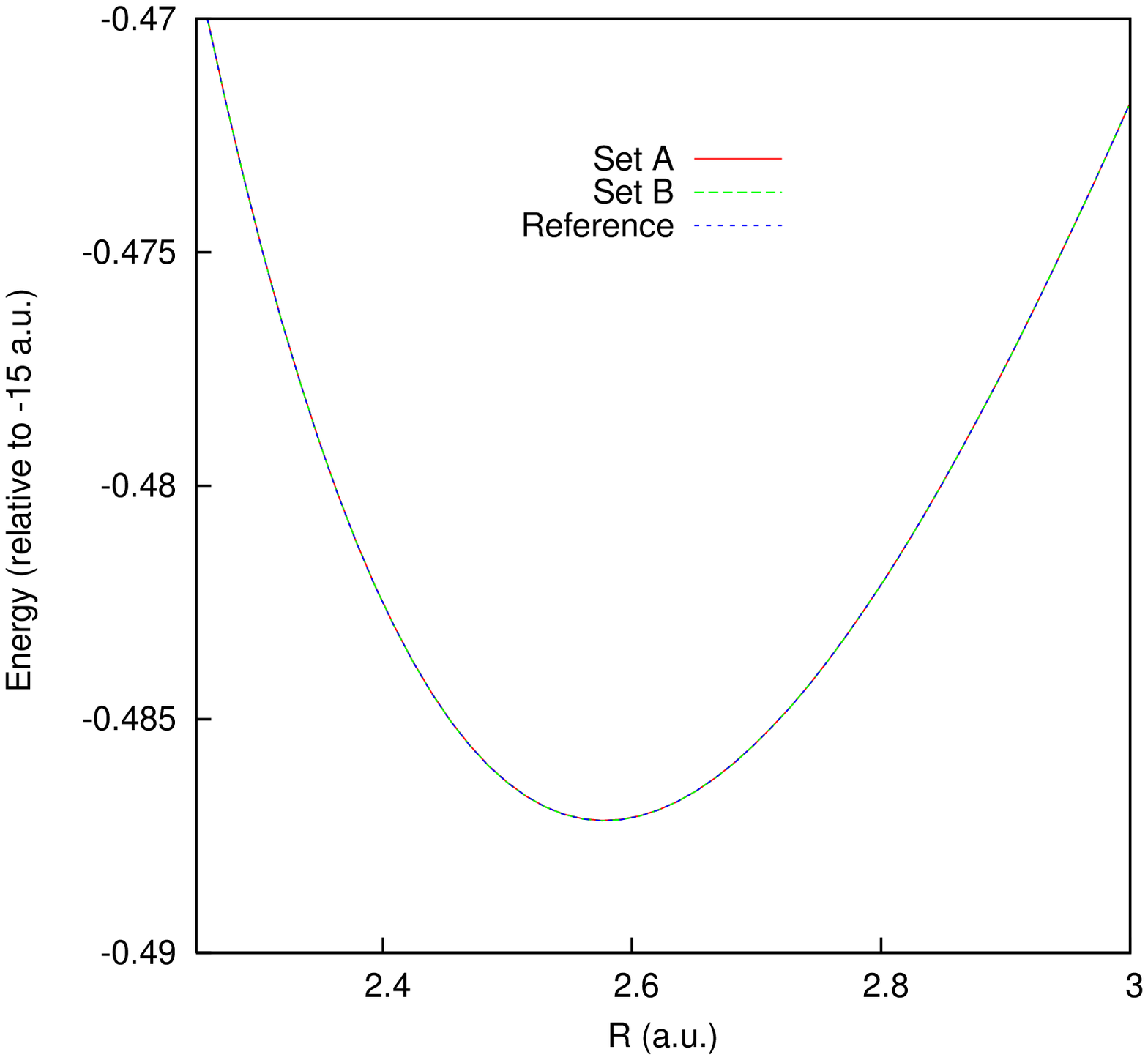}
\end{minipage}%
\caption{Potential energy curves for Cl$_2$ (left panel) and HCl (right panel) for grid Sets A, B. Reference grid-DFT 
results are also given for comparison.}
\end{figure}

For additional test, Table IV reports kinetic, potential and total energies as well as $N$ for 15 species (5 atoms and 10 
molecules) calculated using the BLYP XC functional. With the exception of Mg and S, these are the same species studied in 
Table V of I, using the LDA XC functionals. In this and all other tables henceforth, the experimental geometries
in the NIST database \cite{johnson06} are used. The component energies show similar agreements with the
reference values; hence omitted to avoid crowding. The respective \emph{grid}-DFT results obtained from \cite{schmidt93} are 
presented side by side for comparison. These are ordered in terms of increasing $N$ as descending the table. First 10 of 
these use the same grid parameters as in Table V of I, i.e., $N_r=64, h_r=0.4$, whereas for last 5 we use $N_r=128, h_r=0.3$. 
Overall, the agreement of our results with the reference is excellent. In several occasions (such as Na$_2$, 
P, As, Na$_2$Cl$_2$), the total energies completely match with them. The largest absolute deviation in total energy is 
only 0.0013\% (for NaH). 

\begingroup
\squeezetable
\begin{table}
\caption {\label{tab:table4}Comparison of the kinetic ($\langle T \rangle$), potential ($\langle V \rangle$), total ($E$) 
energies and $N$ for several atoms and molecules with the reference grid-DFT results \cite{schmidt93}. BLYP results in 
a.u. PW=Present Work. See text for details.} 
\begin{ruledtabular}
\begin{tabular}{lrrrrrrrr}
System     & \multicolumn{2}{c}{$\langle T \rangle$} & \multicolumn{2}{c}{$-\langle V \rangle$} & 
\multicolumn{2}{c}{$-\langle E \rangle$}   & \multicolumn{2}{c}{$N$}  \\
\cline{2-3}  \cline{4-5} \cline{6-7} \cline{8-9} 
        & PW   & Ref.~\cite{schmidt93} &  PW & Ref.~\cite{schmidt93} &  PW & Ref.~\cite{schmidt93}  
        & PW   & Ref.~\cite{schmidt93} \\ 
\hline 
Na$_2$        & 0.14723  & 0.14723  &  0.52871 & 0.52871  & 0.38148  & 0.38148  & 1.99999  & 2.00000  \\ 
Mg            & 0.24935  & 0.24935  &  1.06017 & 1.06017  & 0.81082  & 0.81083  & 1.99999  & 1.99999  \\
NaH           & 0.60093  & 0.60088  &  1.34698 & 1.34693  & 0.74606  & 0.74605  & 1.99997  & 1.99999  \\
P             & 2.38891  & 2.38890  &  8.78249 & 8.78248  & 6.39358  & 6.39358  & 4.99999  & 4.99999  \\     
As            & 2.10490  & 2.10494  &  8.14207 & 8.14211  & 6.03717  & 6.03717  & 4.99999  & 4.99999  \\  
S             & 3.73143  & 3.73145  & 13.73598 & 13.73599 & 10.00455 & 10.00454 & 6.00000  & 5.99999  \\
Br            & 4.27022  & 4.27043  & 17.36122 & 17.36148 & 13.09100 & 13.09105 & 7.00000  & 6.99999  \\  
NaCl          & 5.83959  & 5.83957  & 21.01698 & 21.01694 & 15.17739 & 15.17737 & 8.00003  & 7.99999  \\
H$_2$S        & 4.98071  & 4.98066  & 16.21919 & 16.21913 & 11.23848 & 11.23846 & 8.00000  & 7.99999  \\ 
PH$_3$        & 4.18229  & 4.18224  & 12.40101 & 12.40096 & 8.21871  & 8.21872  & 7.99999  & 7.99999  \\  
Br$_2$        & 8.66533  & 8.66527  & 34.88603 & 34.88596 & 26.22070 & 26.22069 & 13.99999 & 14.00000 \\  
H$_2$S$_2$    & 8.88238  & 8.88240  & 30.16535 & 30.16538 & 21.28297 & 21.28298 & 13.99999 & 13.99999 \\
MgCl$_2$      & 11.75947 & 11.75999 & 42.54049 & 42.54103 & 30.78102 & 30.78104 & 16.00004 & 15.99999 \\ 
Na$_2$Cl$_2$  & 11.68815 & 11.68842 & 42.12410 & 42.12438 & 30.43595 & 30.43595 & 16.00002 & 15.99999 \\
SiH$_2$Cl$_2$ & 14.14948 & 14.14945 & 49.04463 & 49.04461 & 34.89515 & 34.89516 & 19.99999 & 20.00000 \\     
\end{tabular}                                                                               
\end{ruledtabular}
\end{table}
\endgroup

Table V displays the calculated negative HOMO energies, $-\epsilon_{\mathrm{HOMO}}$ (in a.u.) and estimated atomization 
energies (in kcals/mole) for the same 12 molecules in Table V of I, using the BLYP XC density functional. These are compared 
with the theoretical \cite{schmidt93} as well the experimental results from the NIST database \cite{afeefy05}. Same grid 
parameters as those in the previous table are employed for these. The $-\epsilon_{\mathrm{HOMO}}$ values completely agree 
with those from the theoretical literature results \cite{schmidt93}, except in two occasions (MgCl$_2$, Na$_2$Cl$_2$), 
where the absolute discrepancy remains only 0.0001 a.u. The calculated atomization energies also show very good agreement 
with the reference theoretical values \cite{schmidt93}. For Na$_2$ HCl, NaCl and H$_2$S, our values are identical as those from 
reference. The largest discrepancy (0.248\%) is observed for Br$_2$. However, both quantities differ significantly from the 
experimental values; but that is a separate issue (not directly related to the main theme of this work). These could possibly
be improved further by employing more appropriate basis functions and/or better XC functionals, and may be considered in 
future works.

\begingroup
\squeezetable
\begin{table}
\caption {\label{tab:table5}Comparison of $-\epsilon_{\mathrm{HOMO}}$(a.u.) and atomization energies (kcals/mole) for 
molecules with the literature data using BLYP XC functional. PW=Present Work. 1 a.u.= 627.509471 kcals/mole. See text for 
details.} 
\begin{ruledtabular}
\begin{tabular}{lrrrrrr}
System    & \multicolumn{3}{c}{$-\epsilon_{\mathrm{HOMO}}$ (a.u.)} 
& \multicolumn{3}{c}{Atomization energy(kcals/mol)} \\
\cline{2-4}  \cline{5-7} 
        & PW  & Theory~\cite{schmidt93} &  Expt.~\cite{afeefy05} &  PW & Theory~\cite{schmidt93}  & Expt.~\cite{afeefy05}  \\ 
\hline
Na$_2$        & 0.1002 & 0.1002 & 0.1798 & 11.51  & 11.51  & 17.0        \\ 
NaH           & 0.1421 & 0.1421 &  ---   & 45.37  & 45.36  & 47.2        \\
HCl           & 0.2785 & 0.2785 & 0.4683 & 89.88  & 89.88  & 102.2       \\   
NaCl          & 0.1733 & 0.1733 & 0.3381 & 88.75  & 88.75  & 97.4        \\
H$_2$S        & 0.2190 & 0.2190 & 0.3843 & 156.59 & 156.59 & 173.2       \\ 
PH$_3$        & 0.2287 & 0.2287 & 0.3627 & 218.72 & 218.73 & 227.1       \\  
Cl$_2$        & 0.2652 & 0.2652 & 0.4219 & 22.89  & 22.90  & 57.2        \\ 
Br$_2$        & 0.2451 & 0.2451 & 0.3865 & 24.28  & 24.22  & 45.4        \\  
H$_2$S$_2$    & 0.2288 & 0.2288 & 0.3418 & 181.66 & 181.68 & 229.6       \\
MgCl$_2$      & 0.2697 & 0.2698 & ---    & 164.04 & 164.07 & 187.4       \\ 
Na$_2$Cl$_2$  & 0.2020 & 0.2021 & ---    & 228.43 & 228.46 & 243.1       \\
SiH$_2$Cl$_2$ & 0.2836 & 0.2836 & 0.4300 & 287.92 & 287.95 & 341.8       \\     
\end{tabular}                                                                               
\end{ruledtabular}
\end{table}
\endgroup

The previous work of \cite{roy08} as well as the ongoing discussion amply demonstrates that the 
approach can produce good-quality results for both LDA and gradient-corrected non-local XC functionals for atoms and 
small-to-medium molecules, within the specifics of basis set. This is further validated in Table VI where 
both LDA and BLYP results are reported for 41 extra molecules to illustrate the broad scope its applicability. The 
experimental geometries are again taken from the NIST database \cite{johnson06}. The kinetic, potential, total energy and 
$N$ are presented. Since it has been clearly established that our obtained results are sufficiently accurate, 
here we omit the reference theoretical values. Only some random checks were made to ensure that this is 
indeed the case. First 29 molecules from top were treated using $N_r=64, h_r=0.4$, while for the remaining 12, we used 
$N_r=128, h_r=0.3$ grid. On the light of all the results and discussion so far, we believe that these represent the correct 
values. 

\begingroup
\squeezetable
\begin{table}
\caption {\label{tab:table6}Kinetic ($\langle T \rangle$), potential ($\langle V \rangle$), total ($E$) energies and $N$ for 
several molecules. LDA and BLYP results are given in a.u. PW=Present Work. See text for details.} 
\begin{ruledtabular}
\begin{tabular}{lrrrrrrrr}
System     & \multicolumn{2}{c}{$\langle T \rangle$} & \multicolumn{2}{c}{$-\langle V \rangle$} & 
\multicolumn{2}{c}{$-\langle E \rangle$}   & \multicolumn{2}{c}{$N$}  \\
\cline{2-3}  \cline{4-5} \cline{6-7} \cline{8-9} 
              & LDA      & BLYP     & LDA      & BLYP     & LDA      & BLYP     & LDA  & BLYP        \\ 
\hline 
Mg$_2$        & 0.50225  & 0.50973  & 2.12547  & 2.13172  & 1.62322  & 1.62199  & 3.99999  & 3.99999  \\ 
MgH$_2$       & 1.32435  & 1.38562  & 3.25978  & 3.35466  & 1.93543  & 1.96904  & 3.99999  & 3.99999  \\
AlH           & 1.21615  & 1.25090  & 3.71761  & 3.77090  & 2.50147  & 2.52000  & 3.99999  & 3.99999  \\ 
SiH           & 1.92640  & 1.96967  & 6.25051  & 6.31591  & 4.32411  & 4.34625  & 4.99999  & 4.99999  \\
AlH$_2$       & 1.78779  & 1.85225  & 4.84983  & 4.94506  & 3.06205  & 3.09281  & 4.99999  & 4.99999  \\
Al$_2$        & 1.35409  & 1.37642  & 5.21624  & 5.24844  & 3.86215  & 3.87203  & 5.99999  & 5.99999  \\ 
PH            & 2.93432  & 2.98889  & 9.90272  & 9.98165  & 6.96840  & 6.99276  & 5.99999  & 5.99999  \\
SiH$_2$       & 2.49908  & 2.56394  & 7.42576  & 7.52084  & 4.92668  & 4.95690  & 5.99999  & 5.99999  \\
SH            & 4.28212  & 4.34594  & 14.86480 & 14.96195 & 10.58268 & 10.61602 & 6.99999  & 6.99999  \\
HSe           & 3.60975  & 3.66865  & 13.32892 & 13.41466 & 9.71917  & 9.74601  & 6.99999  & 6.99999  \\
PH$_2$        & 3.50459  & 3.57825  & 11.07451 & 11.17919 & 7.56992  & 7.60094  & 6.99999  & 6.99999  \\
SiH$_3$       & 3.12838  & 3.21874  & 8.58773  & 8.71139  & 5.45935  & 5.49265  & 6.99999  & 7.00000  \\
HBr           & 4.79802  & 4.86591  & 18.47808 & 18.57517 & 13.68006 & 13.70926 & 7.99999  & 7.99999  \\
MgS           & 4.03663  & 4.08783  & 14.90871 & 14.97037 & 10.87208 & 10.88254 & 8.00000  & 8.00000  \\
NaBr          & 4.40312  & 4.45514  & 17.78349 & 17.85351 & 13.38037 & 13.39837 & 7.99999  & 7.99999  \\
KCl           & 5.80442  & 5.86514  & 20.91611 & 21.00807 & 15.11169 & 15.14294 & 8.00001  & 8.00001  \\
KBr           & 4.42002  & 4.47018  & 17.76799 & 17.83592 & 13.34797 & 13.36573 & 7.99999  & 7.99998  \\ 
H$_2$Se       & 4.18273  & 4.26124  & 14.51438 & 14.62158 & 10.33165 & 10.36034 & 7.99999  & 7.99999  \\
HI            & 3.65476  & 3.71595  & 15.57023 & 15.64787 & 11.91547 & 11.93191 & 7.99999  & 7.99999  \\
SiH$_4$       & 3.65832  & 3.76989  & 9.82333  & 9.97905  & 6.16500  & 6.20916  & 7.99999  & 8.00000  \\
AlS           & 4.47030  & 4.52785  & 16.50709 & 16.57417 & 12.03680 & 12.04632 & 8.99999  & 8.99999  \\
MgCl          & 5.96014  & 6.03442  & 21.70428 & 21.80874 & 15.74414 & 15.77432 & 8.99998  & 8.99999  \\
P$_2$         & 4.80329  & 4.86060  & 17.71797 & 17.77801 & 12.91468 & 12.91741 & 9.99999  & 10.00000 \\
SiS           & 5.17566  & 5.23729  & 19.09644 & 19.16753 & 13.92078 & 13.93024 & 9.99999  & 9.99999  \\
AlCl          & 6.33824  & 6.41277  & 23.24566 & 23.34650 & 16.90741 & 16.93374 & 9.99999  & 9.99999  \\
PS            & 6.19564  & 6.27081  & 22.68040 & 22.77003 & 16.48476 & 16.49922 & 11.00000 & 11.00000 \\ 
S$_2$         & 7.58306  & 7.68138  & 27.62272 & 27.77409 & 20.03966 & 20.09271 & 12.00000 & 12.00000 \\
Se$_2$        & 6.21878  & 6.30169  & 24.55124 & 24.67301 & 18.33246 & 18.37132 & 11.99999 & 11.99999 \\
PCl           & 8.05109  & 8.15428  & 29.34580 & 29.48093 & 21.29471 & 21.32664 & 12.00000 & 12.00000 \\
BrCl          & 9.90905  & 10.03194 & 37.85211 & 38.01782 & 27.94306 & 27.98588 & 14.00000 & 14.00000 \\
SiH$_3$Cl     & 8.82191  & 8.97475  & 29.32193 & 29.52706 & 20.50002 & 20.55231 & 14.00000 & 14.00000 \\
SiCl$_2$      & 12.71813 & 12.87162 & 46.37456 & 46.57075 & 33.65644 & 33.69913 & 17.99999 & 17.99999 \\
S$_3$         & 11.45897 & 11.59948 & 41.56900 & 41.72809 & 30.11003 & 30.12861 & 17.99999 & 17.99999 \\
ClS$_2$       & 13.29292 & 13.45889 & 48.23575 & 48.43954 & 34.94284 & 34.98065 & 19.00000 & 19.00000 \\
P$_4$         & 10.04128 & 10.15505 & 35.88401 & 35.95566 & 25.84274 & 25.80061 & 19.99999 & 19.99999 \\ 
AlCl$_3$      & 17.72370 & 17.94164 & 64.51025 & 64.79488 & 46.78655 & 46.85324 & 23.99999 & 23.99999 \\
S$_2$Cl$_2$   & 19.02565 & 19.26416 & 68.83640 & 69.12631 & 49.81075 & 49.86216 & 25.99999 & 25.99999 \\
PCl$_3$       & 19.50036 & 19.74348 & 70.61349 & 70.91249 & 51.11314 & 51.16902 & 26.00000 & 26.00000 \\
SiHCl$_3$     & 19.09265 & 19.34355 & 68.26422 & 68.57817 & 49.17157 & 49.23462 & 25.99999 & 25.99999 \\
SiCl$_4$      & 24.19691 & 24.50296 & 87.70251 & 88.07579 & 63.50560 & 63.57282 & 32.00000 & 32.00000 \\
PCl$_5$       & 30.92116 & 31.33096 &111.74539 &112.22809 & 80.82423 & 80.89713 & 40.00000 & 40.00000 \\
\end{tabular}                                                                               
\end{ruledtabular}
\end{table}
\endgroup

Finally Table VII compares the calculated $-\epsilon_{\mathrm{HOMO}}$ (in a.u.) and atomization energies (in kcals/mole) with
the literature data, for the same molecules at same geometries as in previous table. Experimental results, wherever available, 
are quoted from \cite{afeefy05}. In this case also, reference results are omitted to avoid crowding. 
But as expected, they show very little deviation from ours. An asterisk in the 
experimental atomization energies denote the values at 298$^{\circ}$K; otherwise they imply 0$^{\circ}$K values. Both LDA and 
BLYP results are given for these quantities side by side. However for the former we also include results obtained with  
another exchange functional for the following reason. It is well-known that the XC potentials derived from the simplest LDA
or generalized gradient approximation (GGA) suffer from improper asymptotic \emph{long-range} behavior. Consequently, 
whereas the ground-state total energies of atoms, molecules, solids can be predicted quite satisfactorily by using these 
functionals, the ionization energies obtained via the HOMO energies (usually off by 30-50\% of the experimental values) as 
well as the excited states are described rather poorly. As mentioned in \cite{roy08}, our primary objective is to extend 
this scheme towards the dynamical studies of atoms and molecules under the influence of strong field such as a laser (through 
multi-photon ionization, high-order harmonic generation and other related phenomena) via TDDFT, that can 
potentially exploit the remarkable developments made in basis-set DFT through many pioneering works over the
years. It is a necessary 
prerequisite that both the ionization potential and higher levels be described more accurately. Recently, the modified 
Leeuwen-Baerends (LB) potential \cite{leeuwen94, schipper00}, $v_{xc\sigma}^{LB\alpha} (\alpha, \beta: \rvec)$, containing two 
empirical parameters have been shown to be quite successful in dealing with the above dynamical situations of atoms, molecules 
(see, for example, \cite{chu05}, and the references therein) as well as for the static property calculations including 
TDDFT-based excited states of molecules. This is conveniently written as, 
\begin{equation}
v_{xc\sigma}^{LB\alpha} (\alpha, \beta: \rvec) = \alpha v_{x\sigma}^{LDA} (\rvec) + v_{c\sigma}^{LDA} (\rvec) +
\frac{\beta x_{\sigma}^2 (\rvec) \rho_{\sigma}^{1/3} (\rvec)} 
{1+3\beta x_{\sigma}(\rvec) \mathrm{ln} \{x_{\sigma}(\rvec)+[x_{\sigma}^2 (\rvec)+1]^{1/2}\}}
\end{equation}
where $\sigma$ signifies up,down spins and the last term containing gradient correction is reminiscent of the popular Becke 
exchange energy density functional \cite{becke88a}, 
$x_{\sigma}(\rvec) = |\nabla \rho_{\sigma}(\rvec)|[\rho_{\sigma}(\rvec)]^{-4/3}$ is a dimensionless quantity, 
$\alpha=1.19, \beta=0.01$. This ensures the desired long-range property, i.e., 
$v_{xc\sigma}^{LB\alpha} (\rvec) \rightarrow -1/r, r \rightarrow \infty.$ 
The HOMO energies, obtained from LBVWN (LB exchange and VWN correlation) combination are presented in column 4. It is 
abundantly clear that LBVWN results are significantly improved over either the LDA or BLYP cases. The LDA ionization energies 
are lower than BLYP values for all the molecules considered and LBVWN values are substantially lower than both these two. 
Evidently in our future work on TDDFT as mentioned above, this feature of LB potential will be highly exploited. Now columns 
6, 7, 8 give the computed LDA, BLYP atomization energies and their experimental analogs. Here also both LDA and BLYP results 
show considerable deviation from the experimental values, which include zero-point vibrational corrections as well as 
relativistic effects. In many cases, LDA results are apparently better than their BLYP counterparts. However this observations 
should not be misconstrued to lead to the conclusion that former is a better candidate than the latter. We note that the 
current work employs the Hay-Wadt-type \emph{ab-initio} effective core potentials which are more suitable for the HF-type 
approaches. Among other factors, use of pseudopotentials and basis sets which are more appropriate for the DFT approaches, may alleviate some of 
the discrepancies encountered here and an undertaking along this direction may be initiated in future. Furthermore, large 
deviations in atomization energies have also been found in other recent DFT works involving all-electron calculations and 
more extended basis sets as well (see for example \cite{cafiero06}). In any case, this is an evolving process and does not 
interfere with the main objective of the present work. Also note that there may be some cancellation of errors in the LDA 
case. Finally note that the extension of this approach to the ``all-electron" atomic, molecular calculations as well as for very 
large systems would be relatively difficult compared to the present pseudopotential case in terms of the grid requirement, because of 
the presence of extra core electrons. Nevertheless reasonable \emph{full} calculation of small to medium molecules are possible. 
This is suggested from some of our preliminary studies in this direction which I am currently engaged into and may be considered
in some future communication.

\begingroup
\squeezetable
\begin{table}
\caption {\label{tab:table7} Negative HOMO energies, $-\epsilon_{\mathrm{HOMO}}$ (in a.u.) and atomization energies 
(kcals/mol) for more molecules. LDA, LBVWN (LB+VWN), BLYP results are compared with experiment \cite{afeefy05}. An asterisk 
indicates 298$^{\circ}$K values. Otherwise, 0$^{\circ}$K results are given. See text for details.} 
\begin{ruledtabular}
\begin{tabular}{lccccccc}
Molecule & \multicolumn{4}{c}{$-\epsilon_{\mathrm{HOMO}}$ (a.u.)} & \multicolumn{3}{c}{Atomization energy (kcals/mol)}  \\
\cline{2-5}  \cline{6-8}  
              &  LDA    & BLYP    &  LBVWN  & Expt.~\cite{afeefy05}    &  LDA     &  BLYP    &    Expt.~\cite{afeefy05}   \\
\hline 
Mg$_2$        & 0.1563  & 0.1530  & 0.2316  &         &  4.04    &  0.22    &    0.9     \\ 
MgH$_2$       & 0.2248  & 0.2221  & 0.3471  &         & 113.75   &  109.09  &            \\ 
AlH           & 0.1741  & 0.1715  & 0.2836  &         &  72.65   &  68.31   &    70.3    \\
SiH           & 0.1647  & 0.1597  & 0.2849  & 0.2900  &  74.47   &  69.59   &    70.4    \\ 
AlH$_2$       & 0.1655  & 0.1631  & 0.2755  &         & 127.69   &  118.90  &    114.7   \\  
Al$_2$        & 0.1407  & 0.1400  & 0.2371  & 0.1984  & 22.92    &  21.42   &    37.0    \\
PH            & 0.2214  & 0.2133  & 0.3519  & 0.3730  & 72.05    &  67.14   &    66.3    \\ 
SiH$_2$       & 0.2056  & 0.2027  & 0.3340  & 0.3278  & 155.85   &  143.93  &    144.1   \\ 
SH            & 0.2229  & 0.2174  & 0.3736  & 0.3830  & 84.83    &   74.85  &    83.9    \\  
HSe           & 0.2117  & 0.2057  & 0.3534  & 0.3618  & 79.44    &  70.16   &            \\ 
PH$_2$        & 0.2170  & 0.2111  & 0.3504  & 0.3610  & 152.77   &  139.92  &    149.2   \\ 
SiH$_3$       & 0.2017  & 0.1969  & 0.3278  & 0.2990  & 193.37   &  171.26  &    212.2   \\ 
HBr           & 0.2688  & 0.2603  & 0.4147  & 0.4292  & 91.60    &  79.11   &    87.5*   \\ 
MgS           & 0.1850  & 0.1766  & 0.2882  &         & 55.90    &  42.15   &    71.7    \\ 
NaBr          & 0.1818  & 0.1729  & 0.3057  & 0.3050  & 87.47    &  78.94   &    86.8*   \\ 
KCl           & 0.1481  & 0.1419  & 0.2805  & 0.3859  & 96.61    &  88.02   &    80.3*   \\  
KBr           & 0.1528  & 0.1449  & 0.2760  & 0.2903  & 87.39    &  79.35   &    69.8*   \\ 
H$_2$Se       & 0.2152  & 0.2075  & 0.3524  & 0.3635  & 167.04   &  146.81  &            \\
HI            & 0.2518  & 0.2432  & 0.3824  & 0.3817  & 82.82    &  72.07   &    45.8*   \\
SiH$_4$       & 0.3188  & 0.3156  & 0.4624  & 0.4042  & 339.43   &  312.02  &    302.6   \\
AlS           & 0.2342  & 0.2240  & 0.3497  & 0.3491  &  93.74   &  77.07   &    88.4    \\ 
MgCl          & 0.1848  & 0.1804  & 0.2750  & 0.2753  &  78.76   &  68.47   &    76.5    \\
P$_2$         & 0.2590  & 0.2526  & 0.3877  & 0.3870  &  96.18   &  81.73   &    116.0   \\
SiS           & 0.2565  & 0.2507  & 0.3828  & 0.3870  & 134.04   &  114.54  &    147.2   \\
AlCl          & 0.2255  & 0.2204  & 0.3383  & 0.3454  & 115.69   &  100.65  &    119.2   \\
PS            & 0.1695  & 0.1645  & 0.3054  & 0.3307  & 81.23    &  63.43   &    105.2   \\
S$_2$         & 0.2007  & 0.2023  & 0.3443  & 0.3438  & 56.75    &  52.47   &    100.8   \\
Se$_2$        & 0.1952  & 0.1951  & 0.3283  & 0.3160  & 58.40    &  54.78   &            \\
PCl           & 0.2093  & 0.2023  & 0.3490  &         & 65.12    &  49.37   &    71.7    \\
BrCl          & 0.2623  & 0.2537  & 0.4133  & 0.4079  & 44.95    &  25.41   &    51.5    \\
SiH$_3$Cl     & 0.2780  & 0.2704  & 0.4317  & 0.4267  & 337.97   & 300.07   &    321.7   \\
SiCl$_2$      & 0.2514  & 0.2448  & 0.3909  & 0.3804  & 190.40   & 155.11   &    202.7   \\
S$_3$         & 0.2392  & 0.2294  & 0.3805  &         & 116.84   & 72.14    &            \\
ClS$_2$       & 0.2225  & 0.2161  & 0.3712  &         & 115.08   & 73.52    &    141.0   \\
P$_4$         & 0.2712  & 0.2575  & 0.3964  & 0.3432  & 200.77   & 142.99   &    285.9   \\
AlCl$_3$      & 0.3081  & 0.2976  & 0.4603  & 0.4414  & 278.02   & 232.88   &    303.4   \\
S$_2$Cl$_2$   & 0.2603  & 0.2499  & 0.4107  & 0.3550  & 151.27   & 90.54    &    192.2   \\
PCl$_3$       & 0.2747  & 0.2660  & 0.4266  & 0.3638  & 189.35   & 133.20   &    229.5   \\
SiHCl$_3$     & 0.3076  & 0.2971  & 0.4632  &         & 335.99   & 273.66   &    361.3   \\
SiCl$_4$      & 0.3194  & 0.3085  & 0.4758  & 0.4333  & 333.91   & 258.60   &    378.6   \\
PCl$_5$       & 0.2825  & 0.2722  & 0.4397  & 0.3748  & 246.22   & 145.33   &    303.2   \\
\end{tabular}                                                                              
\end{ruledtabular}
\end{table}
\endgroup

\section{Concluding remarks}
Pseudopotential density functional calculations were performed for atoms and molecules within the LCGTF framework using CCG
in conjunction with an accurate, efficient Fourier convolution technique to represent the classical Hartree potential in real 
grid. In essence, our previous work (I) on the LDA XC functionals has been extended to test its performance and validity in the 
case of gradient-corrected XC functionals which would be necessary for its further applications to realistic physical 
situations. For this purpose, the widely used BLYP XC functional was chosen. The calculated results of a variety of 
quantities such as energy components, eigenvalues, potential energy curves, ionization energies, atomization energies clearly 
reveal that, they are practically of the same quality as obtained from the available theoretical methods. Furthermore, 
companion LDA and BLYP calculations were performed for a large number of molecules (41) to illustrate its scope of 
applicability for a broad range of systems. Comparison with experiments has been made wherever possible. Additionally, for all
the molecules studied, the LBVWN results show significant improvements in the HOMO energies. This has important relevance to 
our prospective future works on studying real-time dynamics of atoms and/or molecules in a strong laser field. Incorporation 
of other pseudopotentials more suited to DFT as well as more extended and elaborate basis sets would be among some of the important issues which 
may be considered in recent future. More accurate XC functionals could also be employed depending upon the
physical system concerned and the nature of the problem dealt with. Applications to weakly bonded systems, clusters and
of course, to larger systems would further consolidate its success. Finally although one could think of some inherent
errors associated with the incompleteness of the grid, this study confirms that with a judicious choice of the grid
coupled with a correct treatment of the Coulomb potential, these can be reduced to tolerable minima. Thus very satisfactory
results could be obtained.

\begin{acknowledgments}
The project was initiated at Prof. D.~Neuhauser's laboratory at the Univ. of California, Los Angeles, where the core
of the program was written. It was further extended at Prof. S.~I.~Chu's laboratory at the University of Kansas. I
thank them for stimulating discussions and providing the computational facilities. Numerous useful discussions
with Dr.~E.I.~Proynov is gratefully acknowledged. Warm hospitality offered by UCLA and the Univ. of Kansas is greatly
appreciated. An anonymous referee is thanked for valuable constructive comments.
\end{acknowledgments}


\begin{thebibliography}{99}
\bibitem{white89} S.~R.~White, J.~W.~Wilkins and M.~P.~Teter, Phys.~Rev.~B \textbf{39}, 5819 (1989).
\bibitem{chelikowsky94} J.~R.~Chelikowsky, N.~Troullier, K.~Wu and Y.~Saad, Phys.~Rev.~B \textbf{50}, 
11355 (1994).
\bibitem{briggs95} E.~L.~Briggs, D.~J.~Sullivan and J.~Bernholc, Phys.~Rev.~B \textbf{52}, R5471 (1995).
\bibitem{gygi95} F.~Gygi and G.~Galli, Phys.~Rev~B \textbf{52}, R2229 (1995).
\bibitem{beck00} T.~L.~Beck, Rev.~Mod.~Phys.~ \textbf{72}, 1041 (2000).
\bibitem{helgaker00} T.~Helgaker, P.~Jorgensen and J.~Olsen, \emph{Molecular-Electronic Structure Theory}
(John-Wiley \& Sons Ltd., 2000).
\bibitem{payne92} M.~C.~Payne, M.~P.~Teter, D.~C.~Allan, T.~A.~Arias and J.~D.~Joannopoulos, Rev.~Mod.~Phys.~
\textbf{64}, 1045 (1992).
\bibitem{krack00} M.~Krack and M.~Parinello, Phys.~Chem.~Chem.~Phys.~ \textbf{2}, 2105 (2000).
\bibitem{zheng93} Y.~C.~Zheng and J.~Alml\"{o}f, Chem.~Phys.~Lett.~ \textbf{214}, 397 (1993).
\bibitem{becke88} A.~D.~Becke, J.~Chem.~Phys.~ \textbf{88}, 2547 (1988). 
\bibitem{treutler95} O.~Treutler and R.~Ahlrichs, J.~Chem.~Phys.~ \textbf{102}, 346 (1995).
\bibitem{mura96} M.~M.~Mura and P.~J.~Knowles, J.~Chem.~Phys.~ \textbf{104}, 9848 (1996).
\bibitem{murray93} C.~W.~Murray, N.~C.~Handy and G.~J.~Laming, Mol.~Phys.~ \textbf{78}, 997 (1993).
\bibitem{gill93} P.~M.~W.~Gill, B.~G.~Johnson and J.~A.~Pople, Chem.~Phys.~Lett.~ \textbf{209}, 
506 (1993).
\bibitem{lindh01} R.~Lindh, P.~-A.~Malmqvist and L.~Gagliardi, Theor.~Chem.~Acc.~ \textbf{106}, 
178 (2001).
\bibitem{lebedev75} V.~I.~Lebedev and Zh.~Vychisl, Mat.~Mat.~Fiz.~ \textbf{15}, 48 (1975). 
\bibitem{lebedev76} V.~I.~Lebedev and Zh.~Vychisl, Mat.~Mat.~Fiz.~ \textbf{16}, 293 (1976). 
\bibitem{lebedev92} V.~I.~Lebedev and A.~L.~Skorokhodov, Russ.~Acad.~Sci.~Dokl.~Math.~ \textbf{45}, 587 (1992).
\bibitem{boerrigter88} P.~M.~Boerrigter, G.~Te.~Velde and E.~J.~Baerends, Int.~J.~Quant.~Chem.~ \textbf{33}, 87 (1988).
\bibitem{pederson90} M.~R.~Pederson and K.~A.~Jackson, Phys.~Rev.~B \textbf{41}, 7453 (1990).
\bibitem{chen95} X.~Chen, J.-M.~Langlois and W.~A.~Goddard III, Phys.~Rev.~B \textbf{52}, 2348 (1995). 
\bibitem{kong06} J.~Kong, S.~T.~Brown and L.~F\"{u}sti-Moln\'{a}r, J.~Chem.~Phys.~ \textbf{124}, 
094109 (2006).
\bibitem{roy08} A.~K.~Roy, Int.~J.~Quant.~Chem.~ \textbf{108}, 837 (2008).
\bibitem{becke88a} A.~D.~Becke, Phys.~Rev.~A \textbf{38}, 3098 (1988).
\bibitem{lee88} C.~Lee, W.~Yang and R.~G.~Parr, Phys.~Rev.~B \textbf{37}, 785 (1988).
\bibitem{leeuwen94} R.~van Leeuwen and E.~J.~Baerends, Phys.~Rev.~A \textbf{49}, 2421 (1994).
\bibitem{schipper00} P.~R.~T.~Schipper, O.~V.~Gritsenko, S.~J.~A.~van Gisbergen and E.~J.~Baerends, 
J.~Chem.~Phys.~ \textbf{112}, 1344 (20001).
\bibitem{obara86} S.~Obara and A.~Saika, J.~Chem.~Phys.~ \textbf{84}, 3963 (1986).
\bibitem{mcmurchie78} L.~E.~McMurchie and E.~R.~Davidson, J.~Comput.~Phys.~ \textbf{26}, 218 (1978).
\bibitem{schmidt93} M.~W.~Schmidt, K.~K.~Baldridge, J.~A.~Boatz, S.~T.~Elbert, M.~S.~Gordon, J.~H.~Hensen, S.~Koseki, 
N.~Matsunaga, K.~A.~Nguyen, S.~J.~Su, T.~L.~Windus, M.~Dupuis and J.~A.~Montgomery, J.~Comput.~Chem.~ 
\textbf{14}, 1347 (1993).
\bibitem{martyna99} G.~C.~Martyna and M.~E.~Tuckerman, J.~Chem.~Phys.~ \textbf{110}, 2810 (1999).
\bibitem{minary02} P.~Minary, M.~E.~Tuckerman, K.~A.~Pihakari and G.~J.~Martyna, J.~Chem.~Phys.~
\textbf{116}, 5351 (2002). 
\bibitem{miehlich89} B.~Miehlich, A.~Savin, H.~Stoll and H.~Preuss, Chem.~Phys.~Lett.~ \textbf{157}, 200
(1989).
\bibitem{vosko80} S.~H.~Vosko, L.~Wilk and M.~Nusair, Can.~J.~Phys.~ \textbf{58}, 1200 (1980).
\bibitem{pople92} J.~A.~Pople, P.~M.~W.~Gill and B.~G.~Johnson, Chem.~Phys.~Lett.~ \textbf{199}, 557 (1992).
\bibitem{repository} \emph{Density Functional Repository}, Quantum Chemistry Group, CCLRC Daresbury Laboratory,
Daresbury, Cheshire, WA4 4AD, UK.
\bibitem{andzelm92} J.~Andzelm and E.~J.~Wimmer, J.~Chem.~Phys.~ \textbf{96}, 1280 (1992).
\bibitem{baerends73} E.~J.~Baerends, D.~E.~Ellis and P.~Ros, Chem.~Phys.~ \textbf{2}, 41 (1973).
\bibitem{sambe75} H.~Sambe and R.~H.~Felton, J.~Chem.~Phys.~ \textbf{62}, 1122 (1975).
\bibitem{dunlap79} B.~I.~Dunlap, J.~W.~D.~Connolly and J.~R.~Savin, J.~Chem.~Phys.~ \textbf{71}, 4993 (1979).
\bibitem{lapack99} E.~Anderson, Z.~Bai, C.~Bischof, S.~Blackford, J.~Demmel, J.~Dongarra, J.~Du Croz, A.~Greenbaum, 
S.~Hammarling, A.~McKenney, D.~Sorensen, \emph{LAPACK Users' Guide.} 3rd Ed. (SIAM, Philadelphia, 1999). 
\bibitem{johnson06} R.~D.~Johnson III (Ed.), \emph{NIST Computational Chemistry Comparison and Benchmark Database, NIST
standard Reference Database}. Number 101. Release 14, Sept. 2006. (http://srdata.nist.gov/cccbdb).
\bibitem{afeefy05} H.~Y.~Afeefy, J.~E.~Liebman and S.~E.~Stein, ``Neutral Thermochemical Data'' in \emph{NIST Chemistry 
WebBook, NIST Standard Reference Database Number 69}, Eds. P.~J.~Linstrom and W.~G.~Mallard, June 2005, National Institute 
of Standards and Technology, Gaithersburg MD, 20899. (http://webbook.nist.gov).
\bibitem{chu05} S.~I.~Chu, J.~Chem.~Phys.~ \textbf{123}, 062207 (2005).
\bibitem{cafiero06} M.~Cafiero, Chem.~Phys.~Lett.~ \textbf{418}, 126 (2006).
\end{thebibliography}
\end{document}